\documentclass{article}

\author{Claire I. Levaillant\\\\clairelevaillant@yahoo.fr}
\title{Quantum multiple gray scale images encryption scheme in the bit plane representation model}

\usepackage{amsmath, amssymb, amsthm}
\usepackage{epsfig}

\newcommand{\ltm}{\lceil log_2M\rceil}
\newcommand{\m}{\,\text{mod}\,}
\newcommand{\nts}{\negthickspace}

\begin{document}
\maketitle

\begin{center} Abstract \end{center}
Due to the strong correlations among adjacent pixels, bulky data volume, and the handling of various data formatting, traditional encryption schemes are not suitable for image encryption. After introducing a bit-plane quantum representation for a multi-image, we present a novel way to encrypt/decrypt multiple images using a quantum computer. Our encryption scheme is based on a two-stage scrambling of the images and of the bit planes on one hand and of the pixel positions on the other hand, each time using quantum baker maps. The resulting quantum multi-image is then diffused with controlled CNOT gates using a sine chaotification of a two-dimensional Hénon map as well as Chebyshev polynomials. The decryption is processed by operating all the inverse quantum gates in the reverse order. All the secret keys that are used for the quantum encryption process have been previously sent from the sender to the receiver by post-quantum cryptography. The cryptographic scheme can also be used on a classical computer. 

\section{Introduction}

This paper is concerned with encrypting multiple images in a secure way using a quantum computer. With the prospect of having quantum computers becoming a reality, it is important to prepare for new quantum algorithms allowing for a faster processing of the images. So far in the quantum image processing field, multiple image quantum encryption has been little addressed. However, encrypting efficiently multiple images at once using a quantum computer would have nice applications such as the efficient and confidential transmission of medical images or of satellite images used by the military after the latter images have been adequately processed.

Traditionally, a natural way to encode an image is to apply a geometric transformation to it in order to change the positions of the pixels in the image, then modify the pixel values using pseudo-random sequences.
One of the challenges in quantum image encryption is to convert a classical image into an adequate quantum representation using a quantum circuit with the lowest possible complexity. There are two main constraints on this quantum representation. Namely, on one hand, the quantum representation must be suitable for applying a quantum transformation of our choice during the encryption process. On the other hand, at the end of the decryption process, the classical image which is the readable image should be retrieved accurately, instead of probabilistically, using measurements.

With the rapid development of quantum image encryption schemes, the popular traditional image scrambling algorithms such as the Hilbert transform, the Arnold transform and the Fibonacci transform were translated in $2014$ into quantum circuits, see \cite{QITH} and \cite{QITF} respectively. Displaying the pixels in a $2^n\times 2^n$ matrix, the Hilbert scanning matrix is a permutation of the elements of that matrix along the Hilbert curve originally introduced in \cite{HILB}. A quantum implementation was exhibited in \cite{QITH}.
The quantum circuit for the $2D$ Arnold transform got later generalized in \cite{ADRP} to a $2D$ generalized Arnold transform and to a $3D$ generalized Arnold transform in \cite{MQIE} for the sake of scrambling the pixel positions and the images at once in a multiple images quantum encryption scheme. However, in order to do so, the authors need to add the encoding of $2^n-M$ blank images in their quantum multiple image, with $M$ the total number of images to be encrypted and $2^n\times 2^n$ the size of each image. In the present paper we introduce a new multiple images quantum representation so that the number of qubits added corresponding to the blank images is $\lceil log_2L\rceil-\lceil log_2M\rceil$ instead of $n-\lceil log_2M\rceil$ in the case when there are less images than bit planes, where $L$ denotes the total number of bit planes needed to encode the gray values. This novel representation is based on bit planes like in \cite{BAK1} with the difference that it got adapted to quantum multiple images. Also based on \cite{BAK1}, we rather use a quantum Baker transform which has a larger scrambling period than the quantum Arnold transform. We use a two-stage scrambling. First, we do a pixel independent scrambling of the images and the bit planes: for a given pixel, the pixel bit value corresponding to a given image and a given bit plane becomes the pixel bit value associated with a different image and a different bit plane through the permutation operated by the quantum baker map whose partition parameters and iteration parameter depend on the pixel. Second, we do an independent image and independent bit plane scrambling of the pixel positions by using quantum baker maps whose partition parameters and iteration parameter depend on both the image and the bit plane.

Chaotic systems are commonly used as diffusion tools in quantum image encryption methods. The idea to use chaos in encryption algorithms originates in \cite{IEBC}. Chaotic functions have the characteristics of generating completely different sequences through relatively small changes in the tuning parameters or in the initial conditions. One of the best-studied such functions is the logistic function which has been used to model biological populations. However, the parameter range for chaotic behavior is relatively small and for some isolated parameter values, the behavior is not chaotic. A general way to enlarge the parameter range of chaos and in fact make this range infinite is to use a sine chaotification model composed with another given chaotic map, see \cite{SCMO}. This idea was used by the authors in \cite{BAK1}, applied to the $1D$ logistic map. In the present paper we also apply this idea with the base chaotic map being the $2D$ H\'enon map. For each scrambled image, we use a different control parameter, while the two initial conditions depend on the multi-image plaintext. For a given scrambled image, the two pseudo-random sequences obtained, of length the side size of each image, are sorted by increasing order (we impose the elements of each sequence to be distinct with one another). The resulting two pseudo-permutations are then used in the following manner. Each element of the first (resp second) pseudo-random sequence gets applied a Chebyshev polynomial with index issued from the second (resp first) pseudo-random permutation and the resulting two values get multiplied together yielding an element of the interval $[-1,1]$. For each pixel position we thus obtain an element of this interval. We then multiply it by a power of ten depending on the image and take the integer part modulo the $\lceil log_2L\rceil$-th power of two, where $L$ is the number of bit planes. Each integer is then converted into a sequence of bits of length $L$, namely the secret keys, that are ready to be XOR-ed with the bit values of each bit plane.

The paper is organized as follows. After reviewing the main existing representations for quantum images and what was achieved with them in the past, we will introduce a new model which is well-suited to our encryption/decryption scheme. This new representation will in particular allow us to deal with quantum multiple images, instead of simply one quantum image. Then, we will introduce the quantum baker map used for scrambling. The following section will address the chaotification model used for the diffusion stage. It will be based on three different chaotic maps: Chebyshev, H\'enon and the classical Sine. Thus, we will refer to our chaotification model as "CHS chaotification model". We will then present the encryption/decryption scheme.
Further, we will illustrate the scrambling process on some examples and provide explicit quantum circuits for these examples.
Finally, we will discuss the strength of our scheme and what needs to be done for its precise cryptanalysis. Future directions and open questions dealing with bettering the scheme complexity or improving the scrambling process shall also be exposed.

\section{The BRQMI: a bit-plane representation for quantum multiple images}

A list of the main representations for quantum images appears in \cite{SQIR} or \cite{TQIR}. Below, we recall the representations from $2011$, $2013$ and more recently $2018$ and $2022$ upon which our present representation is built.

One of the pioneering models of representation of quantum images was the flexible representation for quantum images, also known as FRQI. It was introduced in $2011$ by the authors of \cite{FRQI}. In the FRQI model, a quantum image appears as a quantum state of the form:
$$\left\lbrace\begin{array}{l}\frac{1}{2^n}\sum_{i=0}^{2^{2n}-1}|c_i>\otimes |i>\\
|c_i>=cos(\theta_i)|0>+sin(\theta_i)|1>\;\text{and}\;\theta_i\in [0,\frac{\pi}{2}]\end{array}\right.$$
where $|i>$ encodes the position in the image and the angle $\theta_i$ encodes the color at this position. For a gray value image with $8$ bit planes, gray values can be converted into angles in the following way:
$$\theta_i=\frac{v_i}{255}\frac{\pi}{2}\in \Big[0,\frac{\pi}{2}\Big],$$
with $v_i$ denoting the pixel value at pixel $i$.
However, with this representation, we have to prepare a large number of the encrypted quantum image in order to obtain the probability amplitude $cos(\theta_i)$ by measurement. There exists some clever way, like in \cite{MRVA} to optimize the number of measurements needed. However, the classical image will still only be retrieved probabilistically instead of accurately.
There exists a complex-valued analog of FRQI that was introduced two years later. The representation reads instead:
$$\frac{1}{2^n}\sum_{j=0}^{2^{2n}-1}(|0>+e^{i\theta_j}|1>)\otimes |j>$$
This representation allows to use a quantum version of the double random phase encryption that was originally introduced by Refregier and Javidi in \cite{DRPE}. The idea in the quantum version is to apply a first controlled random rotation gate on the first qubit, then apply a quantum Fourier transform (QFT) on the next $2n$ qubits, then apply a second controlled random rotation gate on the first qubit again and last apply an inverse QFT on the resulting state. The operation of QFT is defined as:
$$QFT(|j>)=\frac{1}{\sqrt{Q}}\sum_{k=0}^{Q-1}e^{\frac{2i\pi}{Q}jk}|k>$$
This encryption which is the quantum version of the encryption of \cite{DRPE} appears in \cite{QDRP} where the complex valued representation displayed above gets also introduced. The two sequences of $2^{2n}$ random phases serve as secret keys.
Yet another two years later, the authors of \cite{ADRP} enhanced the existing protocol by first operating a quantum generalized Arnold transform to scramble the pixel positions before applying the rest of the protocol. The latter protocol got generalized in \cite{DRPQ} to any scrambling quantum transform but in the reverse order, that is applying the scrambling quantum transform on the outcome of the quantum double random phase encoding.

To remedy the issues of FRQI and of its complex-valued analog, a novel enhanced quantum representation (NEQR) was introduced by the authors of \cite{NEQR}. The advantage of NEQR is that it contains no more probability amplitudes. Moreover, NEQR achieves a quadratic speedup in quantum image preparation. Suppose the range of gray scale values is $2^q$. Then the NEQR representation reads:
$$|I>=\frac{1}{2^n}\sum_{y=0}^{2^n-1}\sum_{x=0}^{2^n-1}|C_0^{yx}\dots C_{q-1}^{yx}>|yx>,$$
where the $C_i^{yx}$ are all bits.
Using this representation, Yu-Guang Yang and all offered a quantum image encryption using one-dimensional quantum cellular automata and showed that their scheme presented in \cite{QCAE}, on top of offering accurate classical image retrieval, has lower computational complexity than the original \cite{QDRP} based on quantum Fourier transform and double phase encoding.

Around the same time, the authors of \cite{QRLP} introduced a similar representation as NEQR in that the gray scale information is encoded similarly, but instead of dealing with cartesian coordinates, they dealt with log-polar coordinates. In log-polar coordinates, every pixel is sampled as $(\rho,\theta)$, where $\rho$ denotes the log-radius and $\theta$ the angular position. The so-called QUALPI representation allowed for affine transformations such as rotation or symmetry, while NEQR did not.

Recently in \cite{BAK1}, Xingbing Liu and Cong Liu offered a new representation for quantum images using less qubits than the NEQR representation. Their model namely encodes each bit plane separately. It is thus named bit plane representation of quantum images (BRQI). Suppose $L$ is the number of bit planes with $2^{k-1}< L\leq 2^{k}$ and so $k=\lceil log_2L\rceil$. Then NEQR uses $2n+L$ qubits. Instead BRQI uses only $2n+\lceil log_2L\rceil$ qubits. Another main advantage of BRQI is that the quantum representation allows for independent bit plane scrambling of the image, while NEQR does not. Considering a usual $8$-bit planes gray scale image, the quantum multiple images will be written as:
$$|I>=\frac{1}{\sqrt{2^{2n+3}}}\sum_{l=0}^{2^3-1}\sum_{x=0}^{2^n-1}\sum_{y=0}^{2^n-1}|P_{lxy}>\otimes |l>|xy>,$$
where $P_{lxy}$ is the bit value in the $l$-th bit plane at pixel $(x,y)$.
We now introduce the new model BRQMI used for our encryption/decryption scheme. Our BRQMI is a combination of BRQI and of the quantum representation model for multiple images QRMMI of \cite{MQIE}. With respect to BRQI, $\lceil log_2 M\rceil$ qubits are added in order to encode the $m$-th image, with $0\leq m\leq M-1$ and $M$ the total number of images to be encrypted. Denoting by $L$ the number of bit planes, the quantum image is the superposition:
$$|I>=\frac{1}{\sqrt{2^{2n+\lceil log_2 L\rceil+\lceil log_2 M\rceil}}}\sum_{m=0}^{2^{\lceil log_2 M\rceil}-1}\sum_{l=0}^{2^{\lceil log_2 L\rceil}-1}\sum_{x=0}^{2^n-1}\sum_{y=0}^{2^n-1}|P_{mlxy}>|mlxy>$$
$P_{mlxy}$ is the bit value in the $l$-th bit plane of the $m$-th image at pixel $(x,y)$. The $0$-th bit plane denotes the least significant bit plane and the $(L-1)$-th bit plane is the most significant bit plane.
With this representation, it is understood that we set:

$$\left\lbrace\begin{array}{cccccc}\text{If}&M<2^{\lceil log_2M\rceil},&\text{then}&P_{mlxy}:=0&\text{when}&M\leq m\leq 2^{\lceil log_2M\rceil}-1\\
\text{if}&L<2^{\lceil log_2L\rceil}, &\text{then}&P_{mlxy}:=0&\text{when}&L\leq l\leq 2^{\lceil log_2L\rceil}-1
\end{array}\right.$$

Moreover, for the sake of our encryption/decryption scheme, because as part of the two stage scrambling process we will apply a quantum baker map in the $(m,l)$ plane, we impose that $M=L$. This will be assumed throughout the rest of the paper. As far as the conversion of the classical multiple images into the quantum multiple images, it does translate in the following way.
On one hand, if $M>L$, that is there are more images than bit planes, then we add more bit planes filled with bit values $0$.
On the other hand, if $M<L$, that is there are less images than bit planes, then we add $L-M$ blank images. It has to be determined in both cases how this would affect the security of the protocol. We note in particular that in the case when there are more bit planes than images, the attacker has no way of deciding on the number of images that are encrypted. This is rather a factor increasing the security of the scheme.

\section{The quantum baker map used for scrambling}

The map is named by analogy with making bread. The original $2D$ baker map introduced by Fridrich in \cite{FDBM} is a chaotic bijection of the unit square onto itself. The baker map divides the unit square vertically on one hand and horizontally on the other hand. The left (resp right) part of the vertical division is mapped to the bottom (resp top) part of the horizontal division. Explicitly, the map is defined by the equations:
$$B(x)=\begin{cases}(2x,\frac{y}{2})& 0\leq x<\frac{1}{2}\\
(2x-1,\frac{y}{2}+\frac{1}{2})&\frac{1}{2}\leq x\leq 1\end{cases}$$
The generalized baker map divides the unit square into $k$ vertical rectangles of length $p_1,\dots,p_k$ and stretches (resp contracts) each rectangle horizontally (vertically) by a factor $\frac{1}{p_i}$ (resp $p_i$). Formally, the map is defined by
$$\forall\,(x,y)\in [F_i,F_i+p_i)\times [0,1),\;B(x,y)=\Bigg(\frac{1}{p_i}(x-F_i),p_iy+F_i\Bigg)$$
with $$\left\lbrace\begin{array}{l}
p_1+\dots+p_k=1\\
F_i=p_1+\dots +p_i\end{array}\right.$$
Then, the chaotic map is discretized to a finite square lattice of points which represent data items. In our framework, the square lattice is the image and the data items are the pixels. The discretized baker map is required to assign a pixel to another pixel in a bijective manner. The discrete baker map is defined for a quantum image of size $2^n\times 2^n$ with

$$2^n=2^{q_1}+\dots +2^{q_k}$$

as
$$\left\lbrace\begin{array}{l}
N_0=0\; \text{and}\; N_i=2^{q_1}+\dots +2^{q_i}\;\text{and}\;
x=N_{i-1},N_{i-1}+1,\dots,N_i-1\\
(x^{'},y^{'})=\Bigg(2^{n-q_i}(x-N_{i-1})+y\,\text{mod}\,2^{n-q_i},N_{i-1}+\frac{y-y\,\text{mod}\,2^{n-q_i}}{2^{n-q_i}}\Bigg)
\end{array}\right.$$

As noticed by the authors of \cite{BAK0}, under some additional conditions on the parameters of the discrete baker map, it is possible to implement the map by swapping qubits. Below, we recall some useful results of \cite{BAK0}.

\newtheorem*{Theorem}{Theorem}
\begin{Theorem} (Due to Hou-Liu-Feng, $2020$ \cite{BAK0}).\\ Let $B=B_{(2^{q_1},\dots,2^{q_k})}$ with $2^n=2^{q_1}+\dots+2^{q_k}$ be a discrete Baker map.\\
If for every integer $i$ with $1\leq i\leq k$ we have $2^{q_i}|2^{q_1}+\dots+2^{q_{i-1}}$, then $B$ has a quantum implementation using quantum SWAP and controlled SWAP gates.\\
Moreover, the number $P_n$ of partitions allowing for quantum implementation for an image of size $2^n\times 2^n$ satisfies to the recursion:
$$P_n=P_{n-1}^2+1\;\text{and}\; P_0=1$$
\end{Theorem}

The proof is based on the following two lemmas, both taken from \cite{BAK0}.

\newtheorem{Lemma}{Lemma}

\begin{Lemma} (cf. $\S\,4.2$ of \cite{BAK0}).\\
(i) The map defined by
$$M_s(x,y)=\Bigg( 2^{n-s}x\,\text{mod}\,2^n+y\,\text{mod}\,2^{n-s},\frac{y-y\,\text{mod}\,2^{n-s}}{2^{n-s}}+x-x\,\text{mod}\,2^s\Bigg)$$
has a quantum implementation. \\
(ii) Each subfunction on $\lbrace N_{i-1},\dots,N_i-1\rbrace$ of the discrete baker map is $M_{q_i}$ if and only if $2^{q_i}|2^{q_1}+\dots+2^{q_{i-1}}$.
\end{Lemma}

\textsc{Proof.} (i) Writing in binary representation $x=x_{n-1}\dots x_0$ and \\$y=y_{n-1}\dots y_0$, we namely have:
$$M_s(x,y)=(x_{s-1}\dots x_1x_0y_{n-s-1}\dots y_1y_0,x_{n-1}\dots x_sy_{n-1}\dots y_{n-s})$$
(ii) Suppose that when $x=N_{i-1},N_{i-1}+1,\dots,N_i-1$, the discrete baker map is identical to $M_{q_i}$. Then,
$N_{i-1}=x-x\,\text{mod}\,2^{q_i}$. This implies that: $$2^{q_i}|N_{i-1}=2^{q_1}+\dots+2^{q_{i-1}}$$
Conversely, suppose the latter condition is satisfied. For temporarily lighter notations, set $n_i:=2^{q_i}$ and $N=2^n$.
Since by assumption we have $n_i|N_{i-1}$, it comes:
$$\frac{N}{n_i}(x-N_{i-1})=\frac{Nx}{n_i}-\frac{N_{i-1}}{n_i}N=\frac{Nx}{n_i}\,\text{mod}\,N$$
It follows that:
\begin{eqnarray*}
N_{i-1}&=&x-(x\,\text{mod}\,2^{q_i})\\
2^{n-q_i}(x-N_{i-1})&=&(2^{n-q_i}x)\;\text{mod}\,2^n
\end{eqnarray*}
\hfill $\square$

The recursion of the theorem providing the number of admissible partitions satisfying to the conditions implying a possible quantum implementation of the discrete baker map relies on the following lemma which is also taken from \cite{BAK0}.
\begin{Lemma} (cf. $\S\,4.3$ of \cite{BAK0}).\\ Suppose $N=2^n=n_1+\dots +n_k$ with $\forall i=1,\dots,k,n_i=2^{q_i}$ and
$n_i|N_{i-1}$ where $N_0=0$ and $N_i=n_1+\dots+n_i$. Then, the following implication holds. \\
$$N_{i-1}\leq \frac{N}{2}\Longrightarrow N_i\leq \frac{N}{2}$$
Consequently, the left and right parts may be partitioned separately and then merged.
\end{Lemma}

\textsc{Proof.} Suppose $N_{i-1}\leq \frac{N}{2}$. The fact that $n_i|N_{i-1}$ and $n_i|\frac{N}{2}$, ensures that
$n_i|\frac{N}{2}-N_{i-1}$. This implies in turn that $\frac{N}{2}-N_{i-1}\geq n_i$ and thus $N_i\leq\frac{N}{2}$\hfill $\square$

For a single image encryption, it appears that the baker map breaks correlation between pixels. For a multi-image encryption,
this fact is even more reinforced if we also permute the images and the bit planes like in the current proposal.
However, changing only one pixel value won't create any drastic change in the ciphertext images. This is an important factor of insecurity. To remedy this, diffusion is needed. It gets discussed in the following section.

\section{The CHS chaotification model used for diffusion}

A common idea for the sake of diffusion is to use a deterministic random bit generator which will be used to change the pixel values of the scrambled quantum image by XOR operations. The bits that are generated by the deterministic random bit generator are pseudo-random in that they are entirely determined by an initial seed. We use some chaotic systems so that any small change in the initial seed will lead to a completely different result. The randomness performance of the chaotic system is measured through the Lyapunov exponent. If $x_{n+1}=f(x_n)$,
$$LE_f=lim_{n\rightarrow \infty}\Big(\frac{1}{n}\sum_{i=0}^{n-1}ln|f^{'}(x_i)|\Big)$$
A positive Lyapunov exponent means a chaotic behavior and larger Lyapunov exponents correspond to higher sensitivity to the initial condition.

One of the classical chaotic systems is the $2$-dimensional H\'enon discrete dynamical system:

$$\left\lbrace\begin{array}{l}
x_{n+1}=1-a\,x_n^2+y_n\\
y_{n+1}=b\,x_n
\end{array}
\right.$$

\noindent It exhibits a chaotic behavior when the two parameters are respectively set to $a=1.4$ and $b=0.3$. It was introduced by Michel H\'enon in the $1970$'s, see \cite{HENO}.

A goal now is to enlarge the Lyapunov exponent and the parameter range of chaos. Then, the sensitivity to the initial condition will be increased and if the range of parameters tentatively becomes infinite, then the key space for the chaotic diffusion model will still be enlarged and hence invulnerable to brute-force attacks.
Start from another chaotic one-dimensional discrete dynamical system, namely
$$x_{n+1}=g(x_n)=sin(\pi x_n),$$
and consider instead the new system
$$x_{n+1}=h(x_n)=sin(\pi\,\lambda\varphi(x_n)),$$
where $\varphi$ is a chaotic map and $\lambda\geq 1$. \\
Observe that:
$$LE_h=LE_g+LE_{\varphi}+ln(\lambda)>LE_{\varphi}$$
The goal is thus achieved and moreover, the parameter $\lambda$ controls the Lyapunov exponent of the new family of chaotic maps. Further, we see that the Lyapunov exponent can be made as large as desired.

We now define a two-dimensional chaotic system:

$$\left\lbrace\begin{array}{l}
x_{i+1}=sin\big(\pi\lambda^{(m)}_1(1-1.4\,x_i^2+y_i)\big)\\\\
y_{i+1}=sin\big(\pi\lambda^{(m)}_2(0.3\,x_i)\big)
\end{array}\right.$$

For each image $m$, we pick different parameters $\lambda^{(m)}_1>1$ and $\lambda^{(m)}_2>1$. The secret keys for the diffusion stage are then obtained in the following manner.

We will run the dynamical system with a seed depending on the total intensity scaled between $0$ and $1$ of the set of images to be encrypted. A total intensity of $0$ corresponds to all white images, while a total intensity of $1$ corresponds to all black images. Thus, we set
$$x_0:=\sum_{\begin{array}{l}\qquad\qquad0\leq x\leq 2^n-1\\\qquad\qquad 0\leq y\leq 2^n-1\\\qquad\qquad 0\leq m\leq M^{'}-1\end{array}}\frac{C_{xym}}{M^{'}(2^L-1)2^{2n}},$$
with $M^{'}$ the total number of images to be encrypted (without counting the blank images) and $C_{xym}$ the pixel value of the $m$-th image at pixel $(x,y)$.
Then we set:
$$y_0:=T_{\lbrace\sum_{x,y,m,l}P_{mlxy}\rbrace}(x_0),$$
where $T_k$ is the $k$-th Chebyshev polynomial, namely the unique polynomial of $\mathbb{R}[X]$ such that:
$$\forall \theta\in\mathbb{R},\,T_k(cos \theta)=cos(k\theta)$$
The polynomial $T_k$ can be expressed as follows:
$$T_k(X)=\sum_{i=0}^{\lfloor k/2\rfloor}\binom{k}{2i}(X^2-1)^iX^{k-2i}$$
The $T_k$'s may also be defined inductively by:
$$\left\lbrace\begin{array}{l}
T_0=1,\,T_1=X\\
T_{k+2}(X)=2XT_{k+1}(X)-T_k(X)
\end{array}\right.$$
With the seed depending on the plaintext image, the diffusion process is not only related to some secret keys, but also related to the plaintext image.
From the seed, iterate the discrete dynamical system as long as is necessary to obtain a sequence of $2^n$ distinct values $x_{m,1},\dots,x_{m,2^n}$ for $x$ and a sequence of $2^n$ distinct values $y_{m,1},\dots,y_{m,2^n}$ for $y$. Assign to the $i$-th horizontal pixel coordinate the $(i+1)$-th value $x_{m,i+1}$ of the pseudo-random sequence for $x$ and to the $j$-th vertical pixel coordinate the $(j+1)$-th value $y_{m,j+1}$ of the pseudo-random sequence for $y$. During the process, ignore the first hundred iterations so as to avoid transient effects. Next, sort the two pseudo-random lists $(x_{m,1},\dots,x_{m,2^n})$ and $(y_{m,1},\dots,y_{m,2^n})$ by increasing order and from there assign to the horizontal $i$-th (resp vertical $j$-th) pixel coordinate the integer $s_{m,i}$ (resp $t_{m,j}$) denoting the index position of $x_{m,i+1}$ (resp $y_{m,j+1}$) in the respective sorted lists. For a pixel $(i,j)$, compute
$$\lfloor T_{s_{m,i}}(y_{m,2^n-i+1})T_{t_{m,j}}(x_{m,2^n-j+1})\times 10^{q_m}\rfloor\;\text{mod}\,2^{2^{\lceil log_2M\rceil}},$$
with for instance $q_m > M$.
From there, obtain a $2^{\lceil log_2M\rceil}$-qubit secret key
$$|K_{2^{\ltm}-1,m,i,j}\dots K_{0,m,i,j}>.$$
The respective secret bit values are used in order to diffuse the respective bit values of the scrambled quantum multiple images by XOR operations, just like described in the next section. 

\section{The encryption/decryption scheme}

We operate a two-stage scrambling, each time using quantum baker maps. First, we do an independent pixel scrambling of the images and of the bit planes. Namely, for each pixel $(x,y)$, we apply a quantum baker map on the two $\lceil log_2M\rceil$-qubits $|m>$ and $|l>$. The parameters, as well as the number of iterations of the QBM depend on the pixel.
We obtain:
\begin{eqnarray*}|I^{'}>&=&\frac{1}{2^{n+\lceil log_2M\rceil}}\sum_{x,y=0}^{2^n-1}\sum_{m,l=0}^{2^{\lceil log_2 M\rceil}-1}|P_{mlxy}>|xy>\,QBM_{x,y}^{r(x,y)}(|ml>)\\
&=&\frac{1}{2^{n+\lceil log_2M\rceil}}\sum_{x,y=0}^{2^n-1}\sum_{m,l=0}^{2^{\lceil log_2M\rceil}-1}
|P^{'}_{mlxy}>|mlxy>
\end{eqnarray*}
We now fix the image and the bit plane and use the quantum baker map to scramble the pixel positions. This time, the parameters and the number of iterations of the QBM depend on the couple $(m,l)$. The expression for the new quantum multi-image after pixel scrambling is:
\begin{eqnarray*}|I^{''}>&=&\frac{1}{2^{n+\lceil log_2M\rceil}}\sum_{x,y=0}^{2^n-1}\sum_{m,l=0}^{2^{\lceil log_2 M\rceil}-1}|P^{'}_{mlxy}>|ml>\,QBM_{m,l}^{\tilde{r}(m,l)}(|xy>)\\
&=&\frac{1}{2^{n+\lceil log_2M\rceil}}\sum_{x,y=0}^{2^n-1}\sum_{m,l=0}^{2^{\lceil log_2M\rceil}-1}
|P^{''}_{mlxy}>|mlxy>
\end{eqnarray*}
The quantum multi-image is now ready for the second main phase of the quantum computation which deals with the independent image diffusion of the pixel values. We use qubit ancillas encoding the secret keys in order to perform
$$2^{2(n+\ltm)}\;\text{CCNOT}\;\text{quantum gates},$$
so as to obtain the quantum multi-image ciphertext:
$$|C>=\frac{1}{2^{n+\lceil log_2M\rceil}}\sum_{x,y=0}^{2^n-1}\sum_{m,l=0}^{2^{\lceil log_2M\rceil}-1}
|K_{m,l,x,y}\oplus P^{''}_{mlxy}>|mlxy>$$

The quantum multi-image ciphertext is then converted into a classical multi-image ciphertext by projective measurements. The later blurred digital images are  transmitted to the receiver. In turn, the receiver converts this classical multiple image back into a quantum multiple image and decrypts them using his or her quantum computer.
The decryption process gets simply achieved by using all the inverse quantum gates in the reverse order in which they were used during the encryption process. A new set of projective measurements is then performed in order to retrieve the original classical images, which are the only information which people can recognize. The possible extra blank images get discarded.

\section{An example of quantum implementation for the scrambling}

Suppose the size of the image is $256\times 256$. By the theorem of $\S\,3$, there are $P_8=1947270476915296449559703445493848930452791205$ admissible partitions for the quantum baker map, one of which is
$(16,8,8,32,64,128)$. This is the partition we chose for our example of quantum implementations of pixel positions scrambling.
Suppose also that there are $8$ images to encrypt at once and the gray values are spread over $8$ bit planes.
Amongst the $P_3=26$ admissible partitions for the bit planes and images mixing, we chose to study the quantum circuit for $(4,2,2)$.

The discrete baker map $B_{(4,2,2)}$ is defined on the integer lattice $[0,2^3-1]\times [0,2^3-1]$ by:

$$B_{(4,2,2)}(x,y)=\begin{cases}
\Big(2x+y\,\text{mod}\,2,\frac{y-y\,\text{mod}\,2}{2}\Big)\!\qquad\qquad\qquad\qquad\qquad\text{if}\; x\leq 2^2-1\\
\Big(2^2(x-2^2)+y\,\text{mod}\,2^2,2^2+\frac{y-y\m 2^2}{2^2}\Big)\;\;\;\;\;\text{if}\;2^2\leq x\leq 1+2^2\\
\Big(2^2(x-2-2^2)+y\m 2^2,2+2^2+\frac{y-y\m 2^2}{2^2}\Big)\,\text{if}\;x\geq 2+2^2
\end{cases}$$

A quick calculation shows that:

$$B_{(4,2,2)}(x,y)=\begin{cases}
(x_1\,x_0\,y_0,\;x_2\,y_2\,y_1)&\text{if}\; x_2=0\\
(x_0\,y_1\,y_0,\;x_2\,x_1\,y_2)&\text{if}\;x_2=1\end{cases}$$

We now introduce two binary functions:

$$\left\lbrace\begin{array}{l}f(x,y)=(x_1\,x_0\,y_0,\;x_2\,y_2\,y_1)\\
g(x,y)=(x_1\,y_0\,x_0,\;y_2\,x_2\,y_1)\end{array}\right.$$

Obviously,
$$B_{(4,2,2)}(x,y)=\begin{cases}f(x,y)&\text{when}\;x_2=0\\
(g\circ f)(x,y)&\text{when}\;x_2=1
\end{cases}$$

We derive the quantum circuit for $B_{(4,2,2)}$:

\begin{center}
\epsfig{file=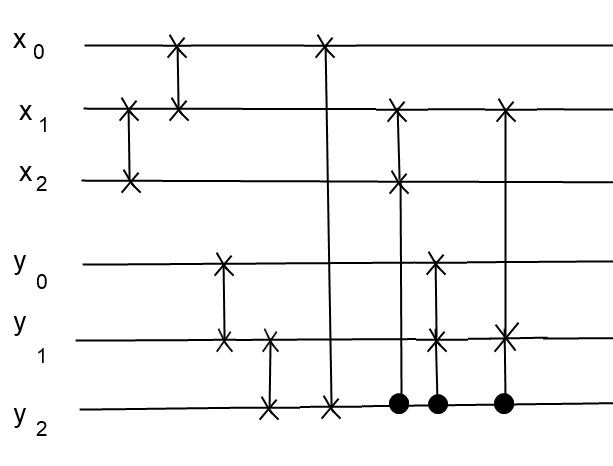, height=6cm}
\end{center}

We will now tackle the quantum circuit for $B=B_{(2^4,2^3,2^3,2^5,2^6,2^7)}$. By using Lemma $1$ of $\S\,3$, we are able to compute the subfunctions like follows:

$$B=\begin{cases}

(x_3x_2x_1x_0y_3y_2y_1y_0,x_7x_6x_5x_4y_7y_6y_5y_4)&\nts\nts\text{when $x_4=x_5=x_6=x_7=0$}\\
(x_2x_1x_0y_4y_3y_2y_1y_0,x_7x_6x_5x_4x_3y_7y_6y_5)&\nts\nts\text{when $x_4=1\,\&\, x_5=x_6=x_7=0$}\\
(x_4x_3x_2x_1x_0y_2y_1y_0,x_7x_6x_5y_7y_6y_5y_4y_3)&\nts\nts\text{when $x_5=1\,\&\, x_6=x_7=0$}\\
(x_5x_4x_3x_2x_1x_0y_1y_0,x_7x_6y_7y_6y_5y_4y_3y_2)&\nts\nts\text{when $x_6=1\,\&\, x_7=0$}\\
(x_6x_5x_4x_3x_2x_1x_0y_0,x_7y_7y_6y_5y_4y_3y_2y_1)&\nts\nts\text{when $x_7=1$}
\end{cases}$$

We have:

$$B(x,y)=\begin{cases}f(x,y)&\text{when $x_4=x_5=x_6=x_7=0$}\\
(g\circ f)(x,y)&\text{when $x_7=1$}\\
(h\circ f)(x,y)&\text{when $x_6=1\,\&\, x_7=0$}\\
(k\circ f)(x,y)&\text{when $x_5=1\,\&\, x_6=x_7=0$}\\
(l\circ f)(x,y)&\text{when $x_4=1\,\&\, x_5=x_6=x_7=0$}\\
\end{cases}$$

with:

$$\left\lbrace\begin{array}{ccc}
f(x,y)&=&(x_3x_2x_1x_0y_3y_2y_1y_0,x_7x_6x_5x_4y_7y_6y_5y_4)\\
g(x,y)&=&(y_6y_5y_4x_7x_6x_5x_4x_0,y_7y_3y_2y_1y_0x_3x_2x_1)\\
h(x,y)&=&(y_5y_4x_7x_6x_5x_4x_1x_0,y_7y_6y_3y_2y_1y_0x_3x_2)\\
k(x,y)&=&(y_4x_7x_6x_5x_4x_2x_1x_0,y_7y_6y_5y_3y_2y_1y_0x_3)\\
l(x,y)&=&(x_6x_5x_4y_0x_3x_2x_1x_0,y_7y_6y_5y_4x_7y_3y_2y_1)
\end{array}\right.$$

\noindent We derive the quantum circuit for this quantum baker map:
\begin{center}
\epsfig{file=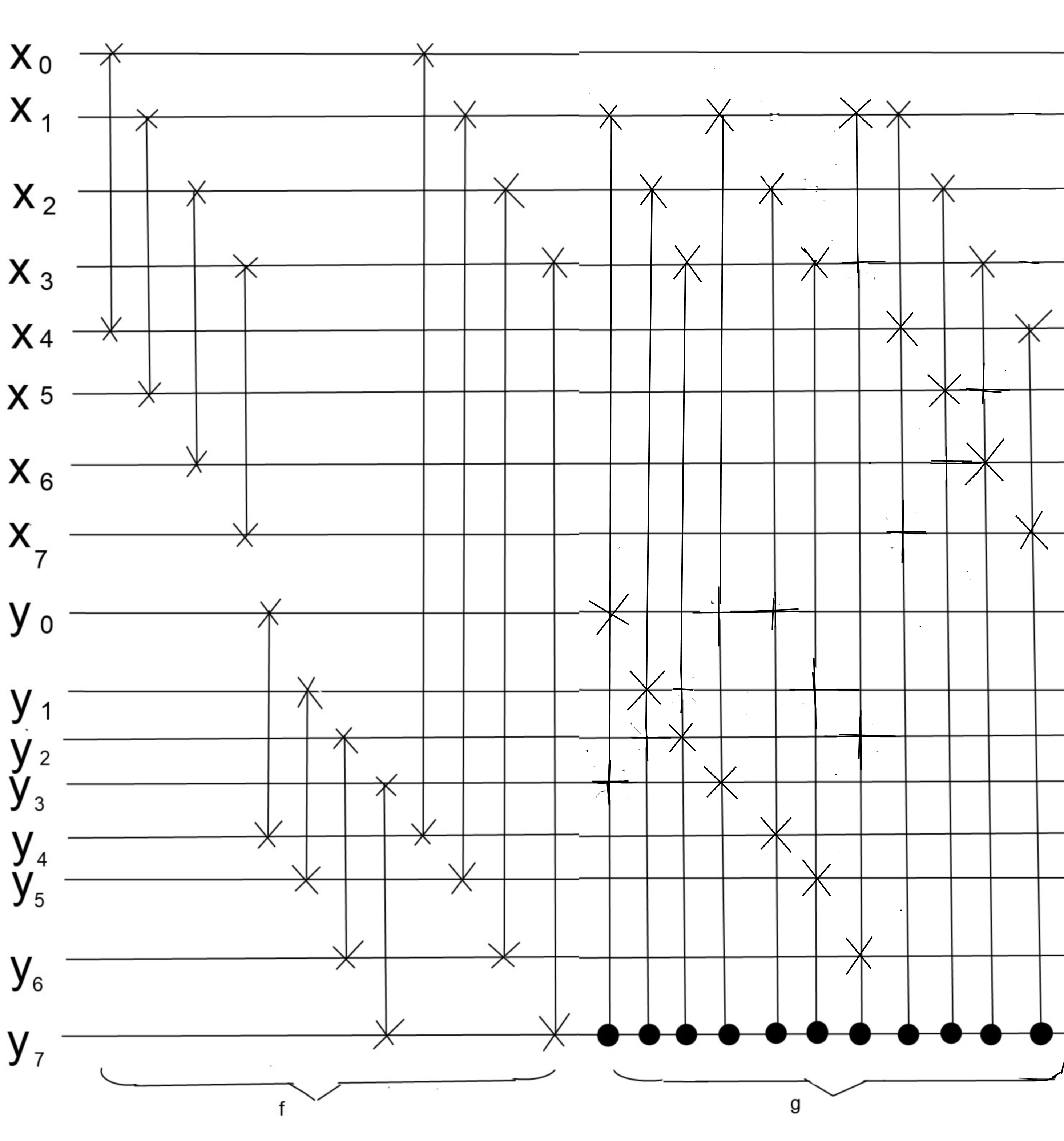, height=8cm}
\end{center}
\noindent Quantum circuit continues:
\begin{center}
\epsfig{file=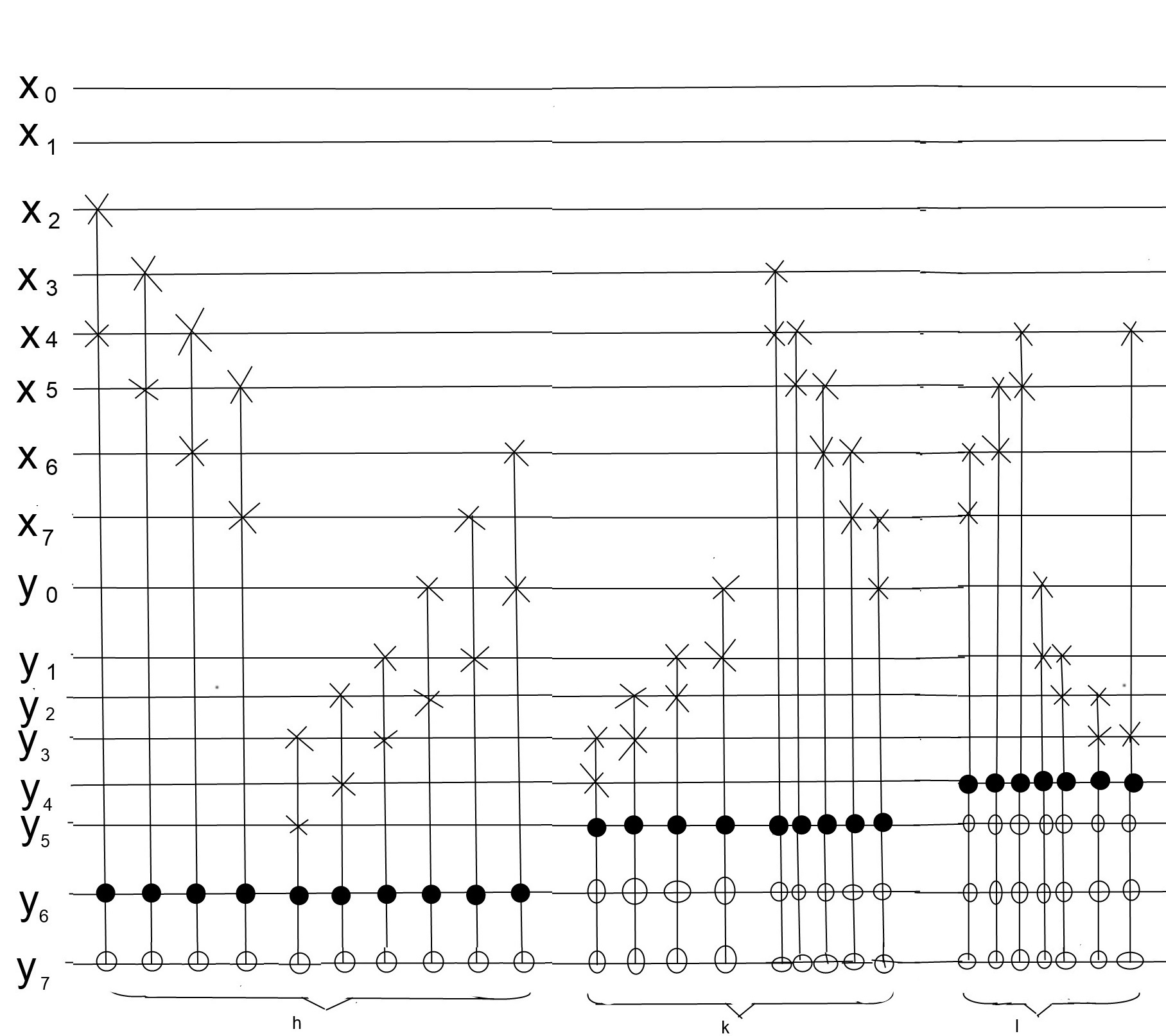, height=8cm}
\end{center}

\section{Conclusion}

Heuristically, the proposed new scheme shows all the signs for robustness. However, the precise cryptanalysis of the sheme is left for future work. The main security tests to be simulated on a classical computer are along the following guidelines.
The distribution of the pixel values can convey some features about the images. Thus, the ciphertext images should hide properly this distribution. For the ciphertext images, a histogram having as horizontal coordinate the gray values and as vertical coordinate the number of pixels with such gray value should have a uniform distribution. Next, the correlation between adjacent pixels should be broken. Also, a subtle difference in the original images should make the ciphertext images completely different. Finally, the main information on the decrypted images should still be obtained in case of occlusion in the ciphertext images (data loss) or interferences in transit (noise).

The efficiency and the overall security of the presented scheme are better compared to single image quantum encryption. Namely, the images get straight away blurred with one another. The pixel correlation gets broken through the use of a different iterated quantum baker map shuffling the bit planes and the images at each pixel. The images from the new set of images get additionally independently scrambled through independent bit plane scrambling of the pixel positions. Each resulting image gets further diffused independently through the use of specific control parameters in the chaotification model. The diffusion is achieved through the use of three different chaotic maps. Using these several chaotic maps has in particular the effect of increasing the sensitivity to the initial conditions and to the control parameters. Since the initial conditions depend on the plaintext multiple image, a slight bit change in one of the plaintext images will completely modify the ciphertext multiple image. The sine chaotification model allows the range of control parameters for chaotic behavior to be $(1,+\infty)$. The key space is thus infinite and invulnerable to brute force attacks. When the number of images to transmit is relatively small, in the sense that this number is less than the number of bit planes, the attackant has no way of deciding on how many plaintext images are transmitted.

In the latter case, our protocol has better storage complexity than the multiple images quantum encryption scheme of \cite{MQIE} as it uses less qubits when it comes to adding some blank images (note also that the BRQI representation of \cite{BAK1} on which our BRQMI representation is based uses less qubits than the NEQR representation of \cite{NEQR} on which the QRMMI representation of \cite{MQIE} is based. Indeed, contrary to the bit plane representation, the NEQR representation uses one qubit for each bit plane). Our scheme is also more general than \cite{MQIE}, where the number of images to be sent cannot exceed the side length of the image. When the number $M$ of images to encrypt is far larger than the number $L$ of bit planes, it would be worth exploring a possible quantum version of the rectangular baker map introduced by Jiri Fridrich in $1997$, see \cite{FDBM}. Using such a map would have the advantage of decreasing the number of qubits. However, it has to be carefully studied how it would affect the complexity of the quantum circuit. In $2004$, Yaobin Mao and his co-authors also generalized the two-dimensional baker map to a three-dimensional one and discretized it. It would be of interest to investigate possible quantum implementations for these discrete three-dimensional baker maps and for the sake of our present scheme generalize them further to four dimensions. On one hand, the number of qubits needed would increase significantly, on the other hand the complexity of the encryption/decryption quantum circuit may be bettered considerably. Finally, it has to be studied whether scrambling in the four directions at once would make the scrambling more efficient or not.\\

\noindent \textbf{Acknowledgement.} The author thanks Salvador E. Venegas-Andraca for introducing her to the field of quantum image encryption.

\newpage

\begin{Large}Appendix A. The H\'enon sine transformation. \end{Large}\\

\begin{center}
\epsfig{file=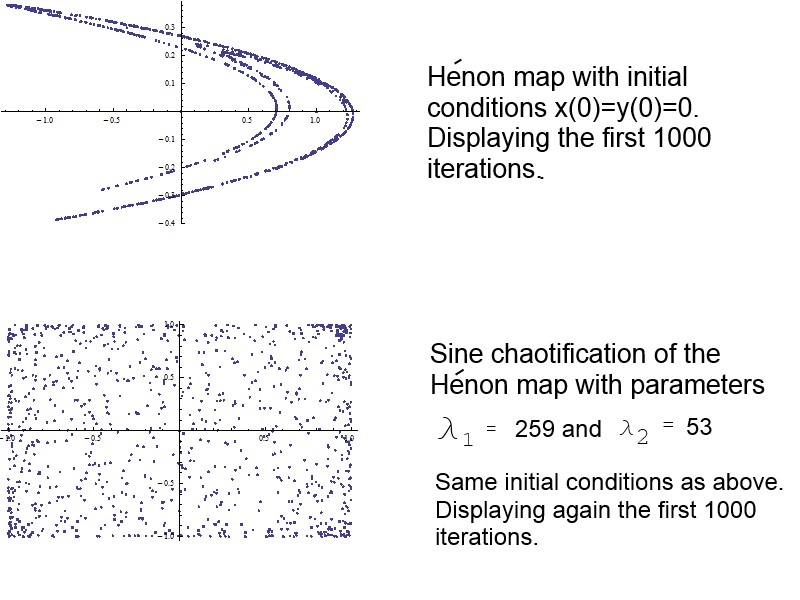, height=10cm}
\end{center}

\newpage
\begin{Large}Appendix B. Chebyshev polynomials.\end{Large}\\
\begin{center}
\epsfig{file=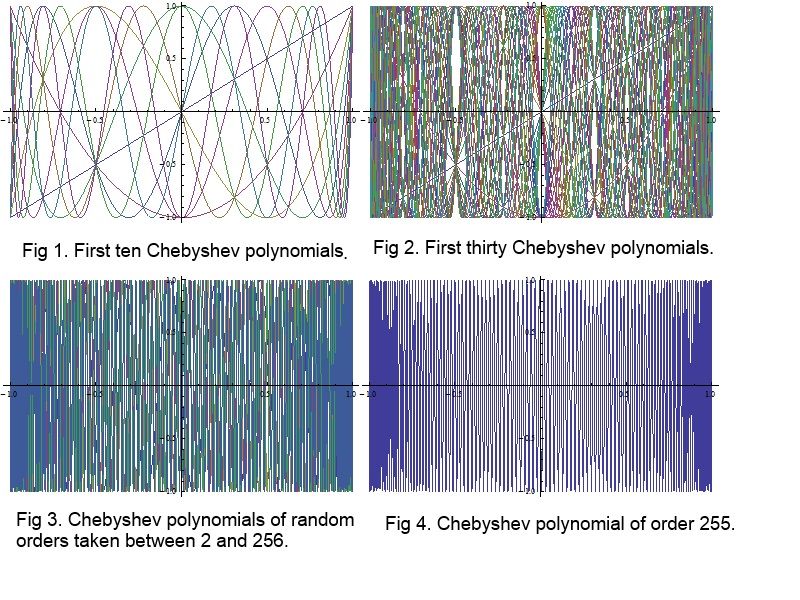, height=10cm}
\end{center}

\end{document}